# Apparatus for dosing liquid water in ultrahigh vacuum

Jan Balajka,[1] Jiri Pavelec,[1] Mojmir Komora,[1,2] Michael Schmid,[1] and Ulrike Diebold[1,*]

[1]Institute of Applied Physics, TU Wien, Wiedner Hauptstraße 8-10/134, 1040 Vienna, Austria.

[2]Central European Institute of Technology, Purkyňova 123, Brno, 612 00, Czech Republic and Institute of Physical Engineering, Brno University of Technology, Technická 2896/2, Brno, 616 69, Czech Republic.

*Corresponding address: diebold@iap.tuwien.ac.at.

**Abstract:**

The structure of the solid-liquid interface often defines function and performance of materials in applications. To study the interface at the atomic scale, we extended an ultrahigh vacuum (UHV) surface-science chamber with an apparatus that allows to bring a sample in contact with ultrapure liquid water without exposure to air. In this process, a sample, typically a single crystal prepared and characterized in UHV, is transferred into a separate, small chamber. This chamber already contains a volume of ultrapure water ice, whose vapor pressure is reduced to UHV range by cooling it to cryogenic temperatures. Upon warming, the ice melts and forms a liquid droplet, which is deposited on the sample. First experiments carried out on rutile $TiO_2(110)$ single crystals using this apparatus exhibit unprecedented purity, as tested by X-ray photoelectron spectroscopy (XPS) and scanning tunneling microscopy (STM). These results enabled to separate the effect of pure water from the effect of low-level impurities present in the air. Other possible uses of the setup are discussed.

1. Introduction

The interaction of water with solid surfaces is the object of interest to a wide variety of scientific fields ranging from meteorology and geochemistry to heterogeneous catalysis, electrochemistry, and solar energy conversion[1]. In many industrial processes, solid surfaces are immersed in aqueous solutions. In an ambient environment, surfaces are covered with a few-monolayer-thick water film formed by condensation from humid air [2]. Whether intentional or not, in practice most surfaces are covered with water.

Although the interaction of water with various classes of materials was studied extensively in the past [1, 3-5], for practical reasons, the vast majority of the research on well-defined surfaces was conducted under UHV conditions. Water does not exist as a liquid in UHV without a background pressure of $H_2O$ vapor. Liquid water placed in a vacuum would evaporate instantly, and larger volumes would freeze as a result of sudden heat loss. In contrast, conventional particle (atom, ion or electron)-based surface-science techniques are restricted to UHV due to the short mean free path in a gas environment. This limitation was overcome in specialized instruments by differential pumping of the analyzer lens system[6] or by using a micrometer-sized liquid jet[7] containing nanoparticles of the material under investigation, e.g., $TiO_2$[8]. Another challenge associated with high pressures is that even low fractions of readily adsorbing impurities become significant within a short time of exposure.

Reports in the literature show examples where the behavior of water on surfaces cannot be extrapolated from UHV to realistic conditions. Kendelewicz *et al.* found a high pressure ($p(H_2O) > 10^{-4}$ Torr) onset threshold for $H_2O$ dissociation on regular $Fe_3O_4(001)$ surface sites [9]. At lower pressures, the hydroxylation was observed only at the defects. In our previous work, we identified a restructuring of $TiO_2(011)$ surface upon exposure to liquid water [10], while low-pressure gas-phase $H_2O$ dosed into a UHV chamber did not alter the surface structure [11, 12].

Here, we present an apparatus designed for dosing ultrapure liquid $H_2O$ on well-defined surfaces of single crystals prepared and characterized in UHV. Similar efforts were made by electrochemical surface scientists, who combined UHV chambers with electrochemical cells to complement electrochemical experiments with *ex-situ* spectroscopic and imaging techniques as well as reliable sample preparation in UHV. Direct transfer from the electrochemical cell to the UHV during or after an electrochemical experiment enabled to acquire snapshots of the actual surface, commonly referred to as emersion experiments[13]. The samples, prepared and characterized in UHV were transferred into a separate compartment and sealed off by pressing the manipulator against a sealing surface[14] or passing the manipulator through differentially pumped O-rings[15, 16]. In systems with exchangeable samples, they could be handed over to another manipulator and the compartment separated from the UHV by a valve. After venting (typically with inert gas) the sample was either transferred through another valve to an external electrochemical cell or the cell itself was inserted into the vented compartment. A frequently used geometry for electrochemical measurements is a hanging meniscus configuration[17-20], where the sample is facing down, and contact with the electrolyte is established only on the one face of the crystal. Other designs work with a flow of the electrolyte across the sample [21, 22].

While there are many variants of electrochemical cells (comprehensive reviews are provided in refs. [13, 23]), the transfer systems are similar and typically constitute a compartment backfilled to an atmospheric pressure of an inert gas before the sample is engaged with an external cell.

In our system, the sample neither leaves the UHV environment nor is it exposed to a venting gas. The sample is transferred in UHV and placed beneath a thick film of ice, grown on a cold tip by vapor transfer from ultrapure liquid $H_2O$ prior to the experiment. While at cryogenic temperature the vapor pressure of $H_2O$ is reduced to the UHV range, by warming-up the tip one can control the vapor pressure evolving from the icicle or eventually allow it to melt and deposit a liquid H₂O droplet onto the sample.

**Experimental setup**

Experiments with liquid water were performed in a custom-built side chamber attached via separately pumped transfer chamber (base pressure $1 \times 10^{-9}$ mbar) to an existing UHV system (base pressure $1 \times 10^{-10}$ mbar). All three chambers were separated by gate valves (see Fig. 1a for a schematic of the UHV system).

a. **Integration within existing UHV-system**

In experiments, the sample is typically prepared in the main chamber (based on an Omicron Compact Lab UHV system) by cycles of sputtering and annealing and can be characterized by various techniques: STM, XPS, low-energy ion scattering (LEIS) and low-energy electron diffraction (LEED). Afterward, the sample is transferred into the small hexagonal chamber (within the red, dashed rectangle in Fig. 1a) where exposure to water takes place. The pressures in the main chamber and the transfer chamber are maintained in UHV range.

Located at the entrance to the transfer chamber is a cryo-panel, made out of an oxygen-free Cu sheet, rolled into a tube, and cooled down by liquid nitrogen (LN₂). This cryo-panel prevents residual water vapor (and other gases condensable at liquid nitrogen temperature) from entering the transfer chamber. In addition, it improves the pressure in the transfer chamber. The tubular shape (32 mm diameter, 176 mm long) allows passage of the magnetic transfer rod that carries the sample.

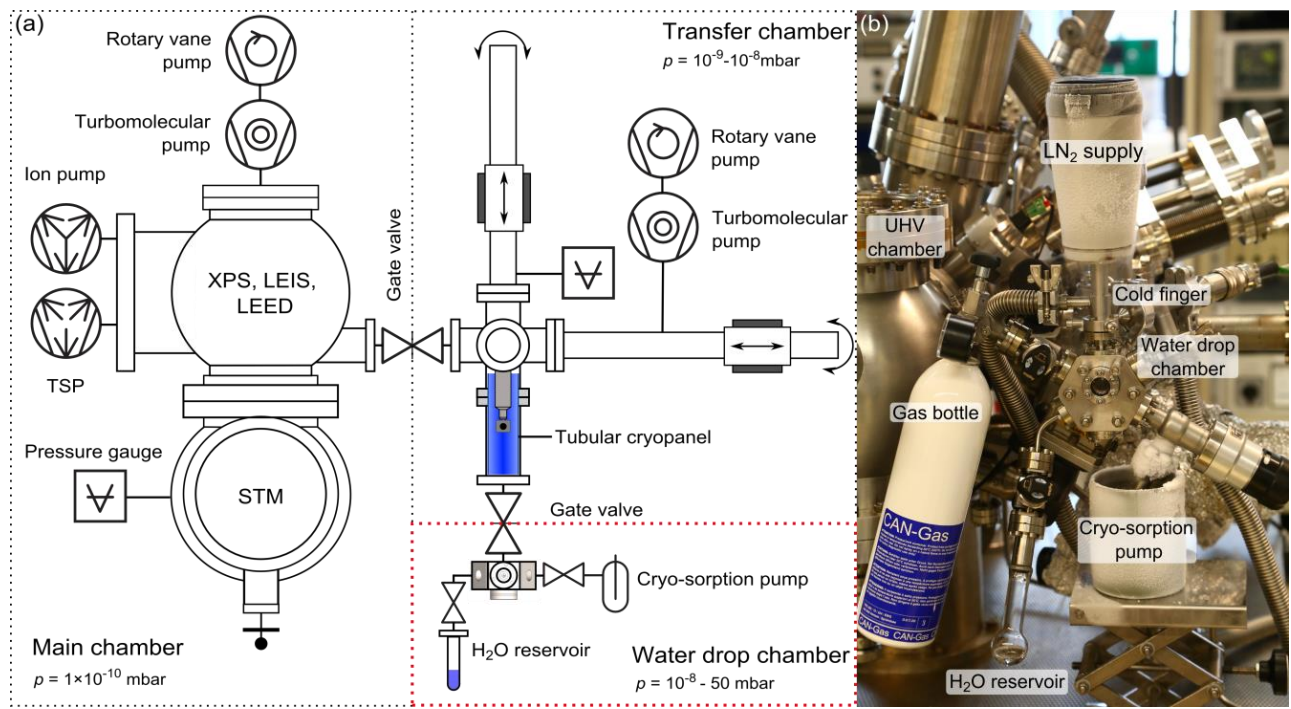

**Fig. 1**. **Overview of the vacuum system and setup for dosing liquid $H_2O$.** (a) Schematic of the UHV system including the hexagonal chamber for water exposure (red, dashed rectangle), (b) photo of the experimental setup. Individual parts of the UHV system in (a) are simplified and not drawn to scale.

### b. Chamber for dosing liquid $H_2O$

A hexagonal shape of the water drop chamber (Fig. 2) was chosen so one could mount the chamber directly onto a gate valve (DN40CF, VAT) of the transfer chamber without a connecting part. This geometry reduces the internal volume and surface area and allows to accommodate up to six DN16CF radial ports. The top, recessed port was used for the so-called cold finger, described in the next section. The two ports at the bottom were used as an inlet for $H_2O$ vapor and outlet to the cryo-sorption pump, respectively. While the inlet $H_2O$ vapor is guided by a tilted hole to obtain an almost direct line of sight to the cold finger, the pumping hole intersects with the volume of the gate valve to achieve better conductance (flow indicated on Fig. 2b). This way, the pumping connects directly to the (dominant) volume of the gate valve and allows for larger hole diameter than would be possible in case of a straight hole into the sample compartment. Located on the front face of the water drop chamber is a DN16CF viewport for observing the samples and the icicle during the transfer and experiment. The drop chamber houses a receptacle for samples mounted on Omicron-type sample plates. The receptacle consist of two symmetric rails mounted to the inner wall of the chamber that guide the sample holder during insertion. Once inserted, the sample rests on a 0.5 mm thin stainless steel wall. A copper rod is pressed to the bottom of the thin wall and allows for sample heating or cooling during the experiment. (Sample heating or cooling was not used in most of the experiments described in the following.) The two additional ports at the top sides were used as an optional gas inlet and another viewport for sample illumination. The internal volume and surface area of the chamber are 25 $cm^3$ and 81 $cm^2$, respectively, and can be reduced to 18 $cm^3$ and 57 $cm^2$ by omitting the two additional ports at the top sides. These numbers do not include the DN40CF gate valve and refer to the chamber with all ports covered by blank flanges excluding the area of blanks. If other parts were mounted to the chamber, this would increase the total surface area and alter the volume.

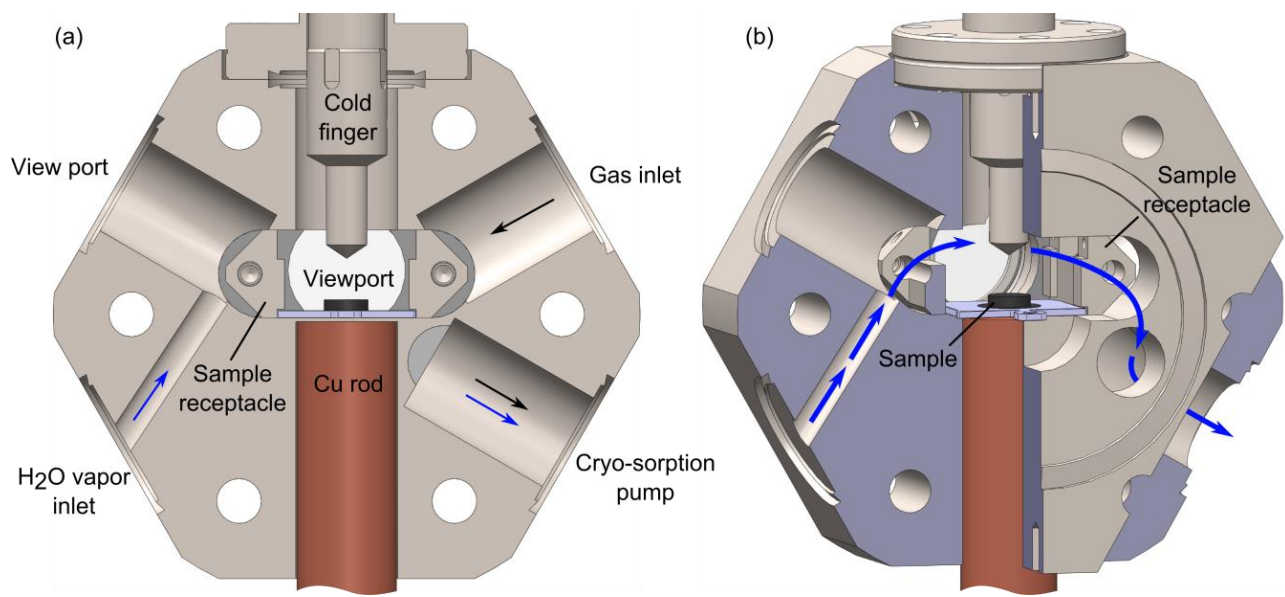

**Fig. 2. Hexagonal chamber viewed from the direction of the transfer chamber:** (a) section-view with a description of ports, (b) perspective partial section-view; blue arrows indicate the diverted flow of $H_2O$ vapor from the supply to the cryo-sorption pump. The flange connecting to the gate valve is at the front and the DN16CF viewport in the back.

### c. Cold finger

The cold finger is a small flow cryostat mounted to the top port of the hexagonal chamber. Its primary function is to create a localized cold spot. When water vapor is introduced into the chamber, it freezes on the cold tip while condensation on other surfaces is avoided. The ice at the tip has to be cooled below 130 K to maintain water vapor pressure below $1\times10^{-10}$ mbar, which is desirable for sample transfer between UHV and the water drop chamber.

Individual parts of the cold finger assembly as referred to in the further text are labeled in Fig. 3. The coolant (here $LN_2$) is guided by a stainless steel capillary (inner diameter 0.5 mm) from the inlet at the top all the way down to the tip. At the tip, $LN_2$ enters a copper heat exchanger pressed onto the inner surface of the tip. $N_2$ expands and returns through a concentric exhaust tube. A particle filter, located at the inlet, prevents (ice) particles from entering the capillary. Circulation of $LN_2$ is achieved by pumping the exhaust with a roughing pump and by the hydrostatic pressure of an $LN_2$ reservoir on top of the inlet. In order to localize the cold spot at the tip of the cold finger, the wider part above the tip is counter-heated (typically to 40°C) by pressing a resistively heated copper tube to the inner stainless steel surface. Heat transfer by conduction between cold and warm areas is minimized by separating them by a thin-wall (0.3 mm) stainless steel tube. To avoid heat transfer by convection, the inner volume of the cold finger is evacuated with a roughing pump. This enables the tip of the cold finger to be locally cooled to cryogenic temperatures while the upper parts are kept above room temperature. The temperatures of both, the cold and warm parts, are measured with K-type thermocouples mounted on the inner surfaces (thermocouples A, B in Fig. 3b). Good thermal contact and a rough vacuum seal between the $N_2$ and the insulation vacuum on the cold side are achieved by pressing the copper parts against the stainless steel surfaces. Defined forces are provided by compression springs. At the top side, the exhaust $LN_2$ is isolated from the rough vacuum by Viton O-rings placed in precision-machined grooves. All parts that reach into the UHV were machined out of stainless steel (SAE 304), while the inner parts were made of stainless steel and copper. Before the final assembly, the outer surface of the tip of the cold finger (reaching into UHV) was boiled and sonicated in pure $H_2O$ as the last step before mounting it to the hexagonal chamber.

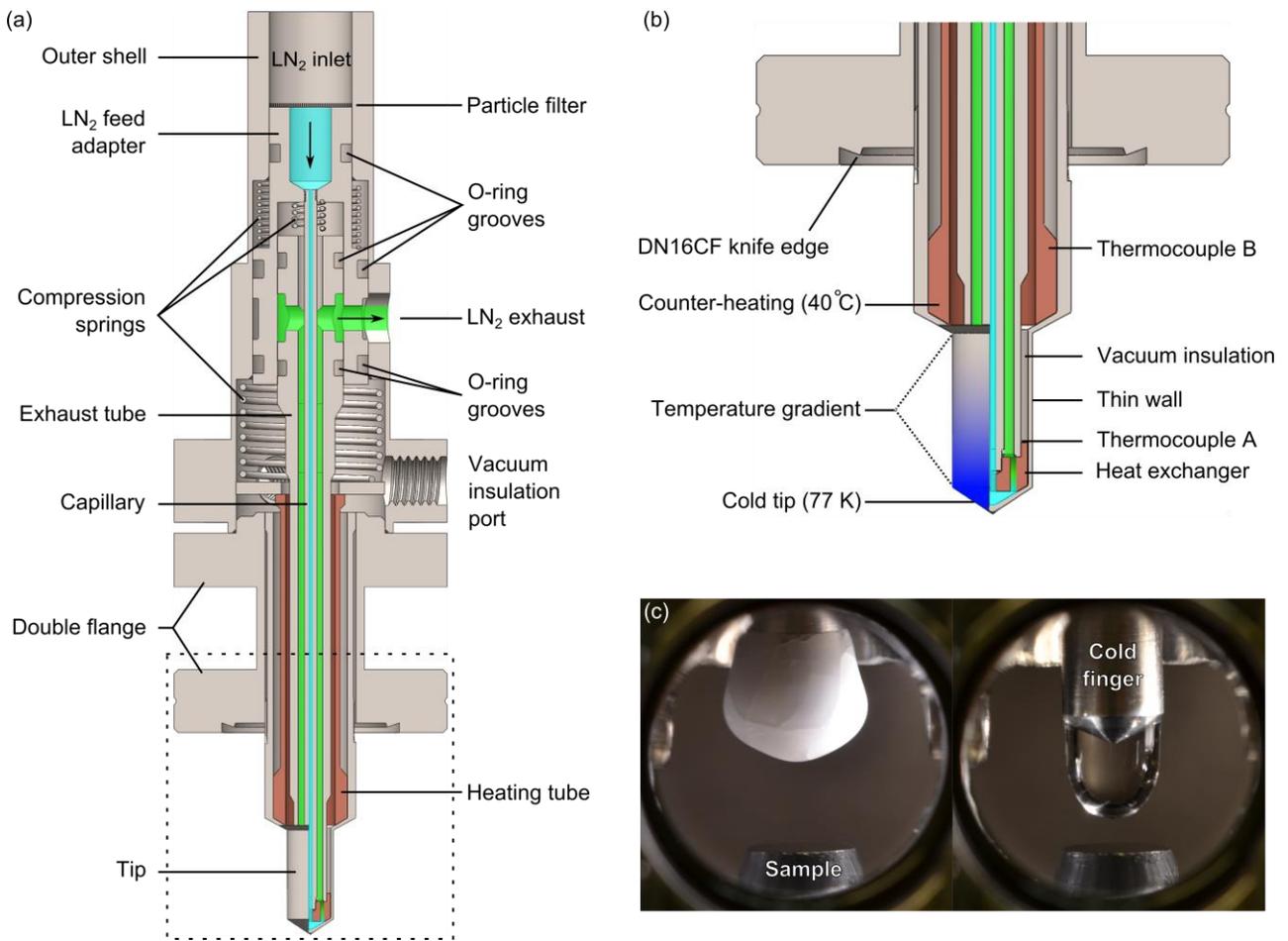

**Fig. 3. Section-view of the cold finger.** The $LN_2$ flowing from the supply to the cold tip is marked in light blue and the exhaust $LN_2$ after passing through the heat exchanger is marked in light green. Parts made of copper and stainless steel are indicated in orange and grey, respectively. The section within the dashed rectangle in (a) is magnified in (b). The blue shading indicates the temperature gradient between the cold

tip and the counter-heated, upper part. The photographs in (c) show ice, locally formed at the lower part of the tip, and a liquid $H_2O$ droplet before deposition onto the sample.

## 2. Experimental procedure

Ultrapure $H_2O$ (MilliQ, Millipore, 18.2 MΩ cm, ≤3 ppb total organic carbon) was used as a water supply. It was placed into a glass vial (see $H_2O$ reservoir in Fig. 1b) that was cleaned in boiling $H_2O$ and extensively rinsed with ultrapure $H_2O$. Afterward, the vial was filled with fresh $H_2O$, mounted to the hexagonal chamber via a valve (Fujikin) and further purified by several freeze-pump-thaw cycles. Before the experiment, a few cycles of connecting the drop chamber to the $H_2O$ reservoir followed by pumping were performed. This procedure cleaned the walls of the drop chamber by removing weakly bound molecules that were replaced by $H_2O$.

The apparatus was then typically used in the following way. The hexagonal drop chamber was evacuated with the turbomolecular pump of the transfer chamber, and the tubular cryopanel at the entrance was cooled down with $LN_2$. The typical pressure in the transfer chamber was $3 \times 10^{-8}$ mbar without a bakeout between experiments. The inner volume of the cold finger and the exhaust were evacuated with roughing pumps to remove residual humidity. Counter heating of the upper parts of the tip was set to 40°C and the heating power regulated by a PID controller. The typical heating power was 10 W (20 V, 0.5 A). The tip of the cold finger was then cooled down, the hexagonal chamber was separated from the transfer chamber by a gate valve, and $H_2O$ vapor was introduced from the reservoir of purified liquid water via an ON-OFF valve (Fujikin). Due to the localized cooling of the cold finger, water froze at the tip where it formed an icicle (see Fig. 3c). The thickness of the ice film, i.e., the volume of the droplet, was determined by the time the valve was opened and by the vapor pressure of the water in the reservoir, which, in turn, was controlled by adjusting its temperature. In practice, the temperature of the bath was stabilized slightly above 0°C to avoid condensation on the drop chamber walls and to allow for controlled growth of icicles.

While the icicle was kept at a cryogenic temperature, a sample was transferred from the main chamber of the UHV system, and placed into the drop chamber, directly under the tip of the cold finger. No pressure increase was observed during the transfer. The drop chamber was then closed off from the transfer chamber, and the cold finger was allowed to warm up by closing the inlet of the $LN_2$ supply. As the tip warmed up, the sample was exposed to the increasing pressure of water vapor evolving from the icicle. Eventually, the icicle melted, and a droplet of liquid water fell on the sample surface. Optionally, additional gas can be introduced to the chamber during this procedure. After appropriate exposure, the hexagonal chamber was pumped with an $LN_2$-cooled sorption pump. We found that additional, efficient pumping could be achieved by cooling the tip of the cold finger, which is favorably located in the center of the hexagonal drop chamber, across from the sample. When an $H_2O$ droplet is pumped from the sample, high pumping speed is not always desirable. In response to a sudden pressure drop, the droplet freezes and the block of ice then slowly sublimes from the sample. A more time-efficient way of pumping is to limit the pumping speed in order to avoid freezing and pump the droplet from its liquid state. After evacuation, the sample was transferred to the UHV chamber for analysis.

## 3. Performance evaluation

The experimental setup described here was already used in our study of liquid water on a prototypical oxide surface $TiO_2(110)$ [24]. There, we clarified the origin of an ordered carboxylic-acid overlayer spontaneously formed on the surface by adsorption or organic acids present in the air. Such overlayers were previously reported by other groups[25-28] after exposure of $TiO_2(110)$ surface to liquid water in the presence of air and incorrectly attributed to ordered interfacial $H_2O$ at the surface [25, 27]. Upon exposure to pure, liquid $H_2O$ using our system, the $TiO_2(110)$ surface retained its original (1×1) structure, and no such overlayer was observed.

In addition, XPS data of the C$1s$ transition are shown here (Fig. 4) to demonstrate the performance of our system. Two procedures are compared. In the first experiment (green curve in Fig. 4), we put a liquid $H_2O$ droplet on a pre-cleaned $TiO_2(110)$ sample inside the transfer chamber after it was vented with high-purity Ar (purity 99.999%, Air Liquide, additionally purified with an inline sorption filter MC50-902 FV from SAES). Outward overpressure flow of argon was maintained during the experiment. Fresh ultrapure $H_2O$ (MilliQ) was deposited onto the sample using a pipet (Eppendorf). The transfer chamber was then evacuated with an $LN_2$-cooled sorption pump (Ultek, Perkin Elmer) for ca 3 minutes before opening the transfer chamber to a turbomolecular pump running at full speed behind a gate valve. The experiment yielded 0.20 ML of carbon evenly distributed between adventitious C (284.3 eV) and carboxylate C (288.3 eV). In the second experiment (blue curve in Fig. 4) a liquid $H_2O$ droplet was deposited on a pre-cleaned sample using the apparatus described here. The latter procedure leads to a visibly cleaner result. After subtraction of the nominally clean UHV-prepared spectrum (black curve in Fig. 4), the remaining carbon coverage is 0.04 ML, almost entirely in the form of adventitious carbon. Moreover, adventitious C is not expected to be homogeneously spread on the surface but rather exists in the form of aggregates (see the STM image in ref. [24]). All coverages are calibrated using a spectrum of saturation coverage (0.5 ML) of formate ($HCOO^-$) on $TiO_2(110)$ (red curve in Fig. 4). It was produced by dosing excess (10 L = $1\times10^{-5}$ Torr s) of formic acid (HCOOH, Sigma Aldrich, purity 98%) in the main UHV chamber. Upon dissociative adsorption of HCOOH, a dense layer of formate forms at a saturation coverage of 0.5 C atom per surface unit cell[29].

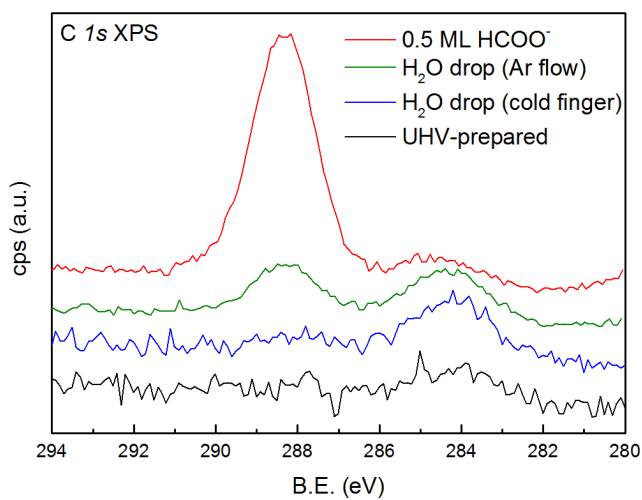

**Fig. 4. Purity evaluation.** Photoelectron spectra for comparison of an ultrapure liquid $H_2O$ drop on the $TiO_2$(110) surface deposited with a pipet in purified Ar flow (carbonaceous contamination 0.20 ML) and liquid $H_2O$ dosed with the system described here (carbonaceous contamination 0.04 ML). Saturation coverage of formate (0.5 ML) gives a quantitative reference. All XPS data were acquired using Mg K$\alpha$ X-rays and a SPECS Phoibos 100 analyzer with a pass energy of 40 eV at grazing emission (70°) from the surface normal. All spectra were normalized to the low-binding-energy background and vertically offset for clarity.

## 4. Potential other uses

The apparatus was designed to prioritize purity from versatility. It offers, however a number of other applications. The system could equally be used for clean dosing of other liquids whose vapor pressure is sufficiently low at $LN_2$ pressures. Alternatively, liquid He could be used as a cooling agent. Isotopically labeled liquids can be dosed, e.g., isotopically labeled $H_2^{18}O$. In a conventional UHV chamber, the $^{16}O$ (from the oxide crystal lattice) and $^{18}O$ (from water) can be distinguished by low-energy ion scattering or by a mass spectrometer upon desorption. An easy modification would be adding more liquid reservoirs, for dosing different liquids. Apart from evaporable liquids, a solution of a solid substance could be prepared directly on a sample. A certain amount of salt would be transferred on the surface, and then diluted with a droplet of pure $H_2O$. Provided that the sample is electrically insulated from the chamber and contacts for other electrodes are included, such a setup could work as a simple electrochemical cell in a droplet. Viewports on the chamber provide good visual access to the sample surface. This makes it possible to measure contact angles of liquids wetting well-defined samples in a well-defined environment similar as in ref. [30]. If UV-grade viewports are used, this setup presents a controlled environment for UV-illumination of samples in contact with liquids and variety of gases. Focused UV-light (or using an aperture to crop the beam) is preferred to illuminate the sample surface only and prevent potential UV-induced desorption from the chamber walls.

A similar setup could be used as a miniature cryostat to locally cool a sample or a part inside a UHV chamber, e.g., to avoid undesirable condensation of gases on other surfaces. Because the cold finger is mounted on DN16CF flange (1.33 in. outer diameter), it offers a compact alternative to commercially available cryostats. Furthermore, the condensation or ice growth of various well-defined substances can be studied in different temperature/pressure regimes and optionally in the background of a variety of gasses. The shape and roughness of the cold surface where condensation occurs could be adapted for that purpose.

## 5. Conclusion

We have presented the design of apparatus for dosing ultrapure liquid $H_2O$ on samples prepared and characterized in UHV. In this closed system, samples are not transferred to an external cell at atmospheric pressure. Instead, a sample is transferred under UHV into a small side chamber and placed in below an icicle of ultrapure water ice, whose vapor pressure is decreased to UHV range by cooling it to cryogenic temperatures. Upon warming, the ice melts, and a liquid $H_2O$ drop is deposited onto the sample. The function of the setup was tested on a $TiO_2(110)$ surface, which proved to be a very sensitive probe to organic impurities. In our previous work, we identified that formic and acetic acids present in air readily adsorb on $TiO_2(110)$ surface[24]. These were not avoided even in high-purity argon under flow conditions as demonstrated by the peak at 288.3 eV (see the green curve in Fig. 4). Using the setup described here, we were able to avoid sample contamination by carboxylic acids, which have a very high affinity to the sample under investigation, and we could substantially reduce the overall level of carbonaceous impurities.

**Acknowledgments:**

This work was supported by the Austrian Science Fund FWF (Wittgenstein Prize project number Z 250-N27 and Doctoral College 'Solids4Fun', project number W1243-N16) and European Research Council (ERC) advanced grant "Oxide Surfaces" (ERC-2011-ADG-20110209). The authors thank Rainer Gärtner and Herbert Schmidt for manufacturing custom-designed components.